\documentclass[fleqn,10pt]{wlscirep}
\usepackage[utf8]{inputenc}
\usepackage[OT1]{fontenc}
\usepackage{color,soul}

\usepackage{siunitx}
\title{Topological invariance in whiteness optimisation}

\author[1,*]{Johannes S. Haataja}
\author[1,2]{Gianni Jacucci}
\author[1,3]{Lukas Schertel}
\author[1]{Silvia Vignolini}
\affil[1]{Department of Chemistry, University of Cambridge, Lensfield
  Road, Cambridge CB2 1EW, UK}
\affil[2]{Laboratoire Kastler Brossel, ENS-PSL Research University, CNRS, Sorbonne Universit\'{e}, Coll\`{e}ge de France, Paris, France}
\affil[3]{Department of Physics, University of Fribourg, Chemin du Mus\'{e}e 3, 1700 Fribourg, Switzerland.}

\affil[*]{jh2225@cam.ac.uk}

\begin{abstract}
  Increasing the light scattering efficiency of nanostructured
  materials is becoming an active field of research both in
  fundamental science and commercial applications. In this context,
  the challenge is to use inexpensive organic materials that come with
  a lower refractive index than currently used mineral nanoparticles,
  which are under increased scrutiny for their toxicity. Although
  several recent investigations have reported different disordered
  systems to optimise light scattering by morphological design, no
  systematic studies comparing and explaining how different
  topological features contribute to optical properties have been
  reported yet. Using in \textit{silico} synthesis and numerical simulations,
  we demonstrate that the reflectance is primarily determined by
  second order statistics. While remaining differences are explained
  by surface area and integrated mean curvature, an equal reflectance
  can be obtained by further tuning the structural anisotropy. Our
  results suggest a topological invariance for light scattering,
  demonstrating that any disordered system can be optimised for
  whiteness.
\end{abstract}

\begin{document}

\flushbottom
\maketitle

\thispagestyle{empty}

\section*{Introduction}
Light diffusion in random structures is a fairly well
established phenomenon that has a profound impact in fundamental
research as well for applications: from random lasing to light 
harvesting to imaging and white paints.\cite{Wiersma2013,Yu2021} 
In fact, bright whiteness is typical of materials 
where light undergoes multiple scattering 
events over a broad range of wavelengths. Such diffusive reflectance
usually requires high refractive index media (with related safety
concerns \cite{EFSA2021}) or thick scattering 
layers to ensure sufficient amount of scattering events. 
Recently, natural examples of disordered materials have drawn much 
attention due to their ability to express record transport 
mean free path using low refractive index components. 
In particular,  the white beetles, e.g. \textit{Lepioda stigma} and 
\textit{Cyphochilus sp. } -- which exploit an anisotropic chitin 
network ($n_c \approx 1.55$) to achieve a bright whiteness -- 
have become the poster children of disordered 
photonics\cite{Vukusic2007,Burresi2014,Wilts2018,Jacucci2019,Lee2020,Lee2021}.
Several investigations have attributed these optical
properties to the specific characteristics of the random 
network structure of the materials. However, due to limited ''metrics''
used to compare and analyse random structures, the role of the various
features of these complex nanostructures in the scattering properties
remains still unclear.

\begin{figure}[ht!]
  \centering
    \includegraphics[width=0.99\textwidth]{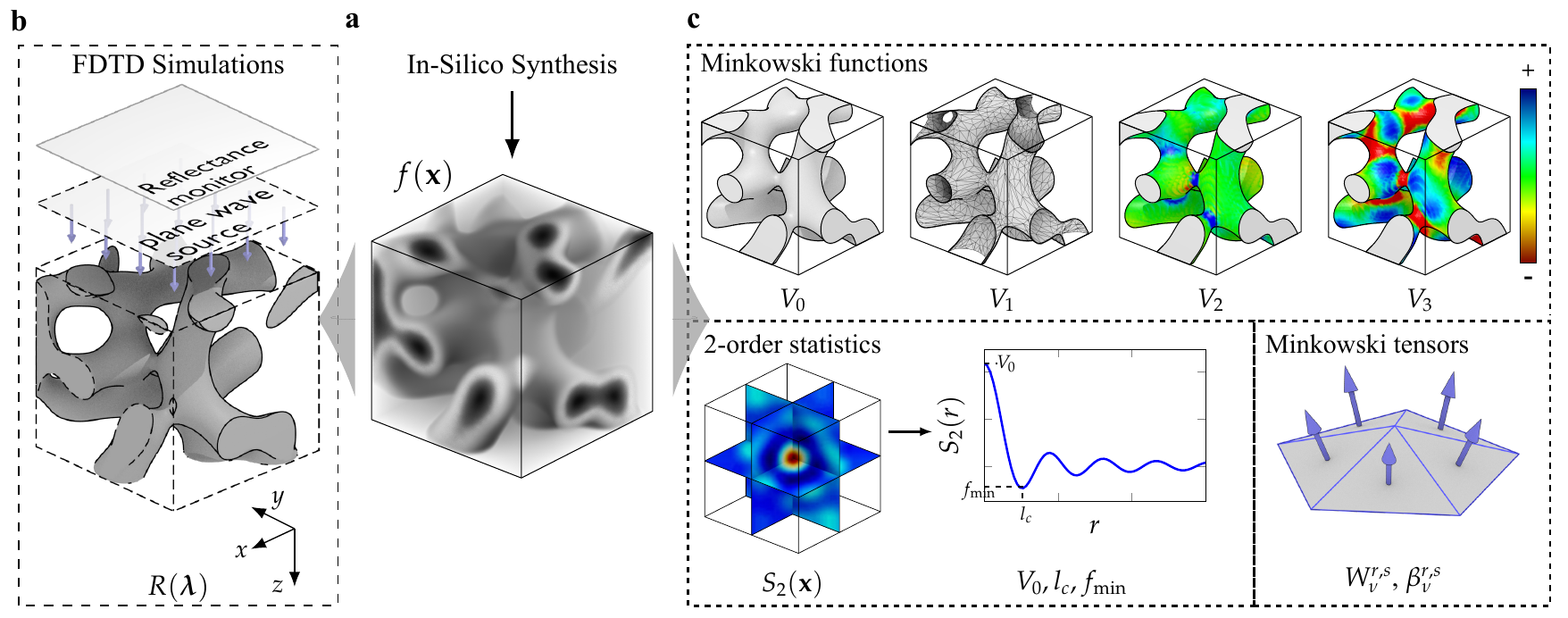}
    \caption{\textbf{Workflow model of the disorder optics
        investigations. a} A disordered structure, $f(\mathbf{x})$, is
      synthesised using stochastic model in silico, and \textbf{b}
      imported to finite difference time domain (FDTD) optics
      simulator with periodic boundary conditions in $x$- and
      $y$-directions and a plane wave propagating along $z$ to obtain
      broad band reflectance spectra,
      $R(\boldsymbol{\lambda})$. \textbf{c} The optical results are
      investigated in respect to structural descriptors which include
      second order statistical measures such as 2-point correlation
      functions $S(r)$, filling fraction $V_0$, and correlation length
      $l_c$, but also more unique structural measures like Minkowski
      functionals, $V_0, V_1, V_2, V_3$ which are the integrated
      volume, surface area, mean and Gaussian curvatures respectively
      (colours represent piecewise contributions), and also Minkowski
      tensors $W^{r,s}_\nu$, and the associated eigenvalue ratios
      $\beta^{r,s}_\nu$ which measure the structural
      anisotropy.\label{fig:intro}}
  \end{figure}

Disordered systems are loosely defined as inhomogeneous materials that
exhibit small (but not long range) structural correlations at relevant
length distances $\mathbf{r}$. The most popular measures used to
characterise disordered systems in optics are based on first and
second order statistics such as form- and structure factors, 2-point
correlation function $S_2(\mathbf{r)}$, filling fraction,
$V_0 (:=S_2(0))$, and correlation length $l_c$. It has been suggested
that unique structural similarities between disordered systems can be
inferred from identical reflectance spectra $R(\lambda)$ and
$S_2(r)$\cite{Burg2019}. Yet in the case of random network 
structures inspired by white beetles, many different
synthetic models have been reported to match the beetle structures
using these functionals
\cite{Meiers2018,Utel2019,Burg2019,Zou2019}, suggesting its
non-uniqueness. This is not surprising, given that in the field of
disordered studies, which predates the study of disordered photonics,
it is well known that only in few exceptions, e.g. in the case
Gaussian random fields, also known as Gaussian Processes (GP), the
structural properties of stochastic fields are uniquely determined by
second order statics alone \cite{Rasmussen2006}, and for an arbitrary
ones more robust topological measures are needed for unique
characterisation. Therefore we propose to use a particularly
useful metrics for this quantification: the Minkowski 
functionals $V_0,V_1,V_2,V_3$, which are based on surface integrals, 
and are proportional to the filling fraction, surface area, 
integrated mean and total curvature respectively \cite{Arns2001,Mantz2008}.

Using such topological identifiers we tested 
the scattering efficiency of a series of disordered structures 
synthesised \textit{in silico} using both top-down and bottom-up
models to cover a wide range of topological features in terms of 
porosity, connectedness, branching and length scale.
We quantified these differences using the Minkowski functionals 
and measured the reflection properties using 
\textit{finite-difference time domain} (FDTD)
simulations (cf. Fig. \ref{fig:intro}). Then, by correlating the 
two, we were able to understand how 
each feature contributes to overall scattering properties, demonstrating
 that disordered  systems are sufficiently explained by second order 
 statistics alone. Finally using the the tensorial Minkowski
measures\cite{Schroeder-Turk2011,Schroeder-Turk2013}, 
we also quantified structural anisotropy and investigated its 
role in whiteness optimisation.

\section*{Results}
\subsection*{In silico synthesis of disordered structures}
\begin{figure}[h!]
  \centering
    \includegraphics[width=0.99\textwidth]{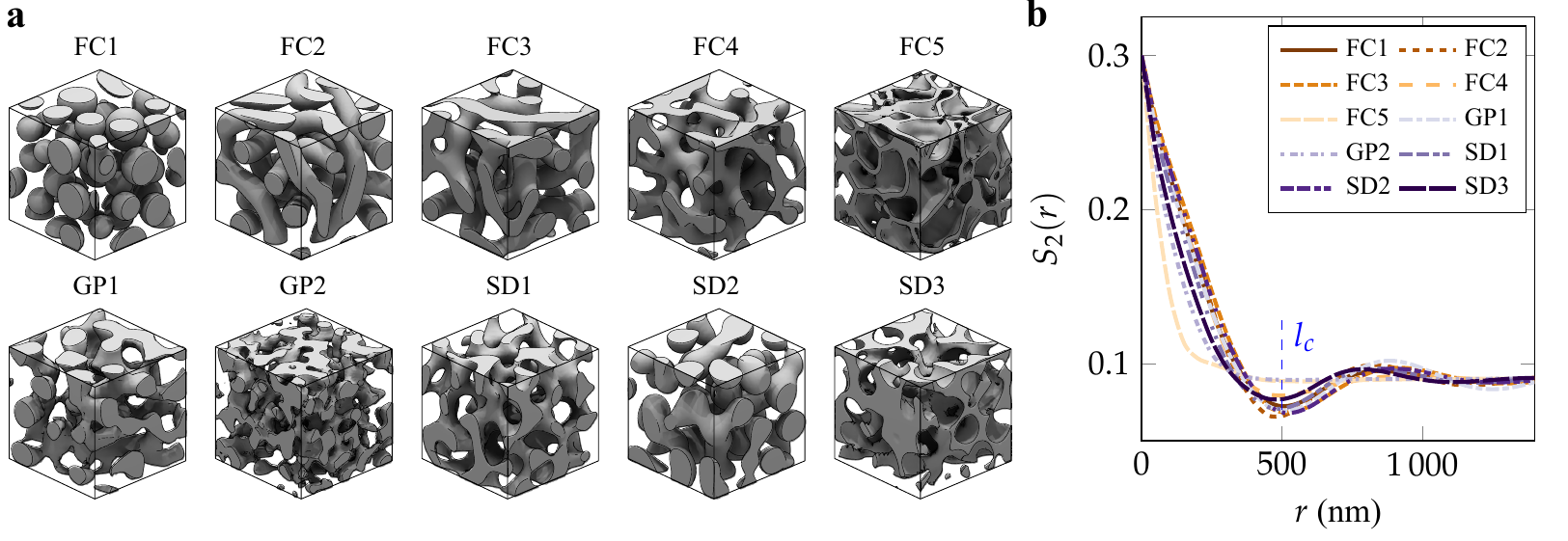}
    \caption{\textbf{a} Investigated simulated structures morphologies
      including Functionalised Cahn-Hilliard (FC*), Gaussian Processes
      (GP*), and spinodal decomposition (SD*) using the Cahn-Hilliard
      model. \textbf{b} The second order correlation functions, $S_2(r)$, 
      of these structures demonstrating that very different topologies
      are often indistinguishable using second order statistics
      alone. \label{fig:morph}}
\end{figure}

As the size of the parameter space for disordered structures
is enormous, we selected 10 different model systems (GP1-2, FC1-5, 
 and SD1-3, cf. Figure \ref{fig:morph}a) using three different stochastic
approaches and mapped the reflectivity landscape using line searches by
varying i) the correlation length $l_c$, ii) filling fraction $V_0$ and
anisotropy of these systems. We believe that our selection 
covers a significant wide range of different topological features
relevant to the topic of scattering optimisations.

In the first approach we used Gaussian Processes (GPs), which are
completely defined by their mean $m(\mathbf{r})$ and $S_2(\mathbf{r})$
(the latter is also known as covariance kernel in machine learning
literature), to test the dependency of photonic properties of
disordered structures on second order statistic alone. We selected two
different correlation functions, one with sinc-type and other with
squared exponential

\begin{equation}
  \text{GP1: }\ \  S_2(r) \propto  \frac{\sin{(lr)}}{lr},  \hspace{15mm} \
  \text{GP2: }\ \ S_2(r) \propto \exp
  \left(
    \frac{-r^2}{2l^2} \label{eq:GP}
\right)
\end{equation}
where $l$ is a scaling parameter, to investigate the effect of a fixed
correlation length, (GP1), vs.~one with distribution of correlation
lengths (GP2) (cf. Figure \ref{fig:morph}a), on the optical response.

In the second case, we used a popular phase-field approach, the
Cahn-Hilliard (CH) model \cite{Cahn1958} to generate more complex
structures. The CH model successfully describes the time
evolution of demixing process (known as spinodal decomposition) of
homogeneous mixture into separate domains and was chosen as it
has recently been suggested
as the formation mechanism behind the beetle scale structures
\cite{Burg2019}. We used three spinodal decomposition (SD) models,
which we initialised using \SI{50}{\percent}, (SD1),
\SI{30}{\percent}, (SD2), and \SI{70}{\percent}, (SD3) filling
fractions.

Although the CH model has been successfully used to simulate
disordered structures in many different fields, it's mainly limited to
bicontinuous networks and isolated micelle morphologies, that exclude
the possibility to investigate e.g. tubular and cellular
morphologies. A particularly interesting approach, that was developed
to gain control over these features, is the Functionalised
Cahn-Hilliard (FCH) model \cite{Gavish2012,Jones2012,Jones2013} which
takes into account hydrophobic and mixing entropy effects, allow to
control the surface curvature. Therefore in the third approach we
adopted a known FCH protocol with \SI{20}{\percent} initial filling
fraction\cite{Gavish2012}, to simulate a variety of structures from
colloidal, (FC1), and tubular, (FC2), to branched and cellular
(FC3-FC5) ones). These structures are also interesting as they offer
flexibility in visual and quantitative (using the $V_n$) matching of
synthetic disordered systems with biological ones.

\subsection*{FDTD simulations}
\begin{figure}[h!]
  \centering
  \includegraphics[width=0.99\textwidth]{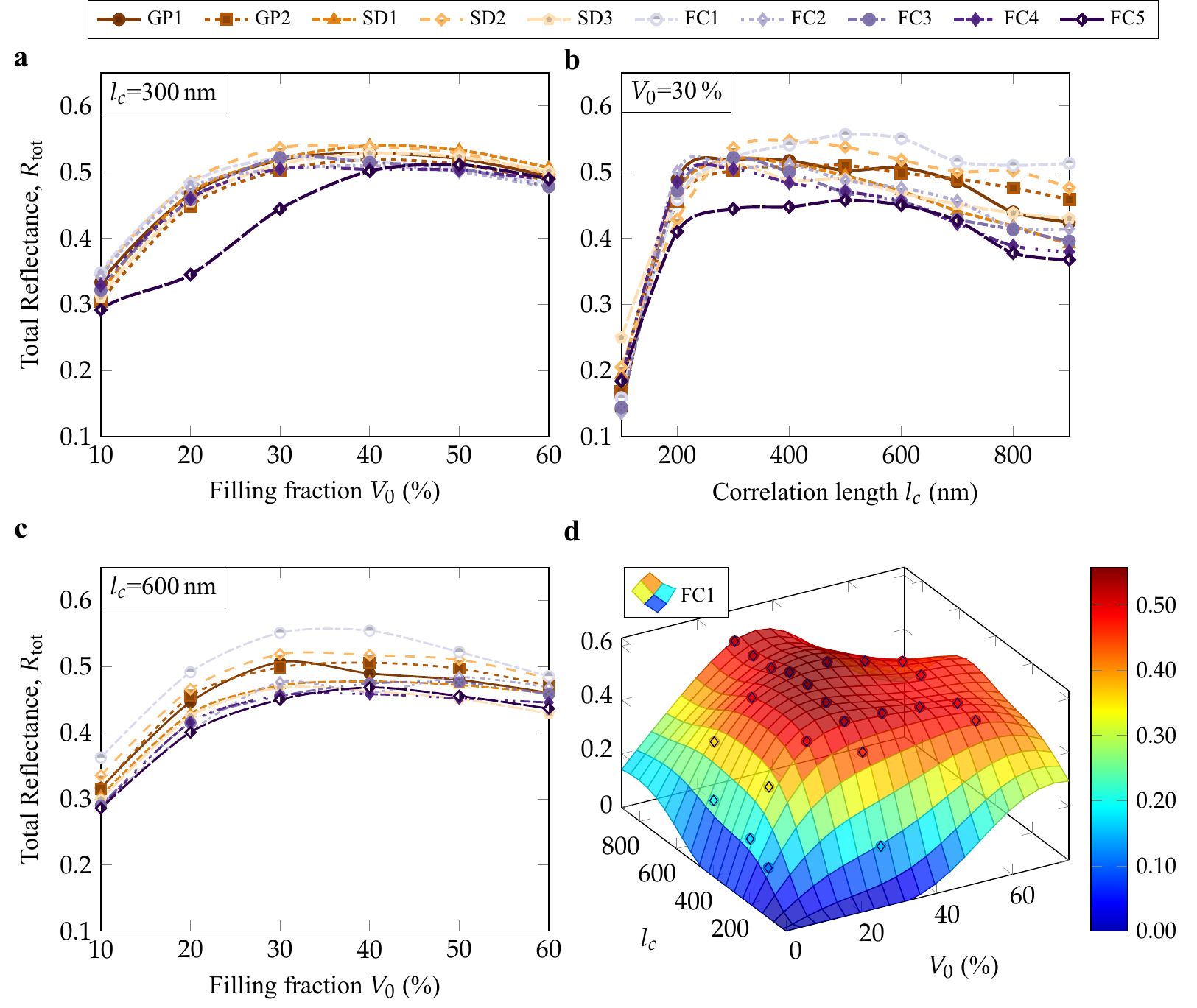}
  \caption{Line searches along different filling fractions $V_0$ 
  and correlation lengths $l_c$ for different morphologies. 
  \textbf{a,c} Varying filling fraction with fixed $l_c = \SI{300}{nm}$
  and   $l_c = \SI{600}{nm}$. \textbf{b} Varying correlation 
  length with fixed $V_0 = \SI{30}{\percent}$. \textbf{d)} 
  Extrapolated surface for the most reflective system FC1. \label{fig:res}}.
\end{figure}

FDTD simulations were carried out using a broadband plane wave
source and a reflectance monitor above the sample. The reflectance
spectra were then integrated over the monitor area, and spectrally 
averaged to obtain total reflectance $R_{\text{tot}}$.  Because the 
parameter space, spanned by correlation length $l_c$ and $V_0$, is 
rather large, and the FDTD simulations are computationally relatively 
costly, we carried out series of line scans by keeping the $l_c$ 
fixed and varying $V_0$, and vice versa, for each model.

Starting from $l_c =\SI{300}{nm}$ and $V_0=30$\% one would
have expected (under the assumption that optical properties are not solely 
determined by 2nd order statistics) to see clear difference in 
reflectivity between different structures. Surprisingly not only were the
average reflectances between different structures, very similar, 
from 0.49 to 0.53, except for the cellular structure FC5 with $R\approx
0.45$, but they all behaved in a similar unimodal manner as function of
$V_0 = [10,20,30,40,50,60]$\%, as shown in Figure \ref{fig:res}a,
with highest reflectivity reached between
$V_0=$\SIrange{30}{50}{\percent} in agreement with earlier
studies\cite{Pattelli2018}. Next, we therefore
continue to scan over correlation
lengths between $l_c=[\SI{100}{nm},\SI{200}{nm},\dots,\SI{900}{nm}]$
while keeping the $V_0=\SI{30}{\percent}$ fixed, cf. Figure
\ref{fig:res}b. Here it could be seen that the value
$l_c=\SI{300}{nm}$ was a convergence
point for most structures. When moving to higher 
values, the 
differences became more pronounced and in favour of colloidal systems,
like FC1 and SD2, and interestingly at the expense
of inverse/cellular structures, such as FC5 and SD3. We then line scanned
$V_0=[10,20,30,40,50,60]$\% only to discover similar unimodal
behaviour and an optimal region of $V_0=$\SIrange{30}{40}{\percent},
with the exception of more pronounced separation in averaged reflection
between the different structures.

\subsection*{Role of anisotropy}
\begin{figure}[h!]
  \centering
  \includegraphics[width=0.65\textwidth]{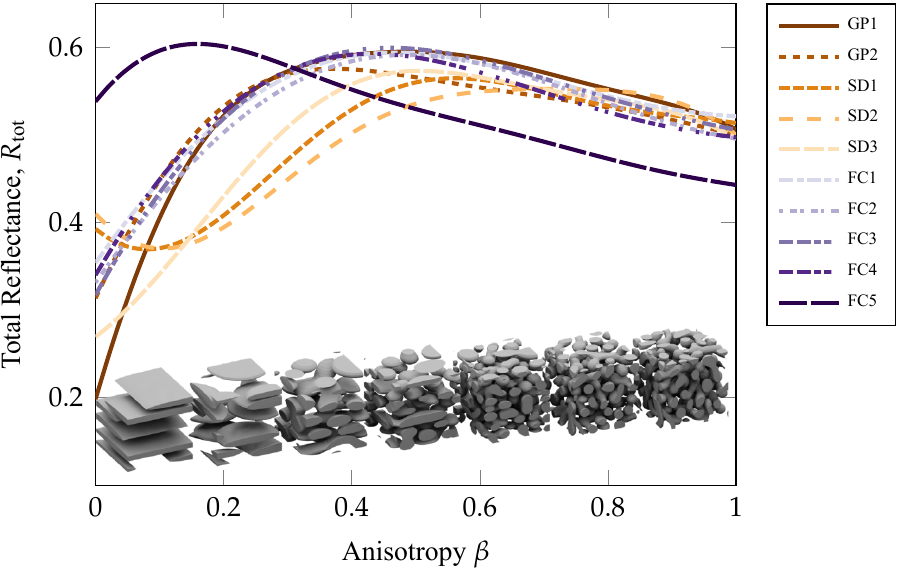}
  \caption{Effect of anisotropy $\beta$ on average
    reflectance. A value of  $\beta=1$ indicates complete
    structural isotropy and $\beta=0$ full structural anisotropy.
    The lines present Gaussian Process fits to measured (FDTD
    simulated) reflectances. The data points have been omitted for 
    clarity, and can be found ESI, Figure S3. Snapshots of FC1
    structures are shown in the bottom to illustrate the
    effect of the anisotropy on a morphology. 
    \label{fig:aniso}}
\end{figure}

As structural anisotropy is known to be an important
feature affecting the reflectance properties of disordered
systems\cite{Jacucci2019b}, we decided to use the Minkowski
tensors, which are based on strong mathematical foundation of integral
and convex geometry\cite{Schroeder-Turk2011}, to
  quantitatively correlate the structural anisotropy to the optical
  response. In fact, to the best of our knowledge, a robust way to
quantify the anisotropy of an arbitrary structure\cite{Jacucci2019b}
is still lacking in the photonics community. Popular methods
include the simple comparison of ratios of correlation
lengths along different directions, but question how such values
should be interpreted remains open, and can be rather inaccurate for
latent anisotropy, as we will later demonstrate.

In 3D there are several different linearly-independent tensors, and 
a particularly suitable one for two-phase structures is

\begin{align}
  W^{0,2}_1(I(\mathbf{x})) := \frac{1}{3}\int_{\partial I} \mathbf{n} \otimes \mathbf{n}\  dA
\end{align}

which measures the distribution of surface normals $\mathbf{n}$
\cite{Schroeder-Turk2010,Schroeder-Turk2013}. The degree of anisotropy
can then be expressed as the ratio of minimal and maximal eigenvalue
of the tensor

\begin{align}
  \beta^{0,2}_1:=\frac{|\mu_{\min{}}|}{|\mu_{\max{}}|} \in [0,1]
\end{align}

which we will now on referred as $\beta$. An isotropic
structure will have $\beta \approx 1$ and lower values signify
increased anisotropy.

\begin{figure}[h!]
  \centering
  \includegraphics[width=0.99\textwidth]{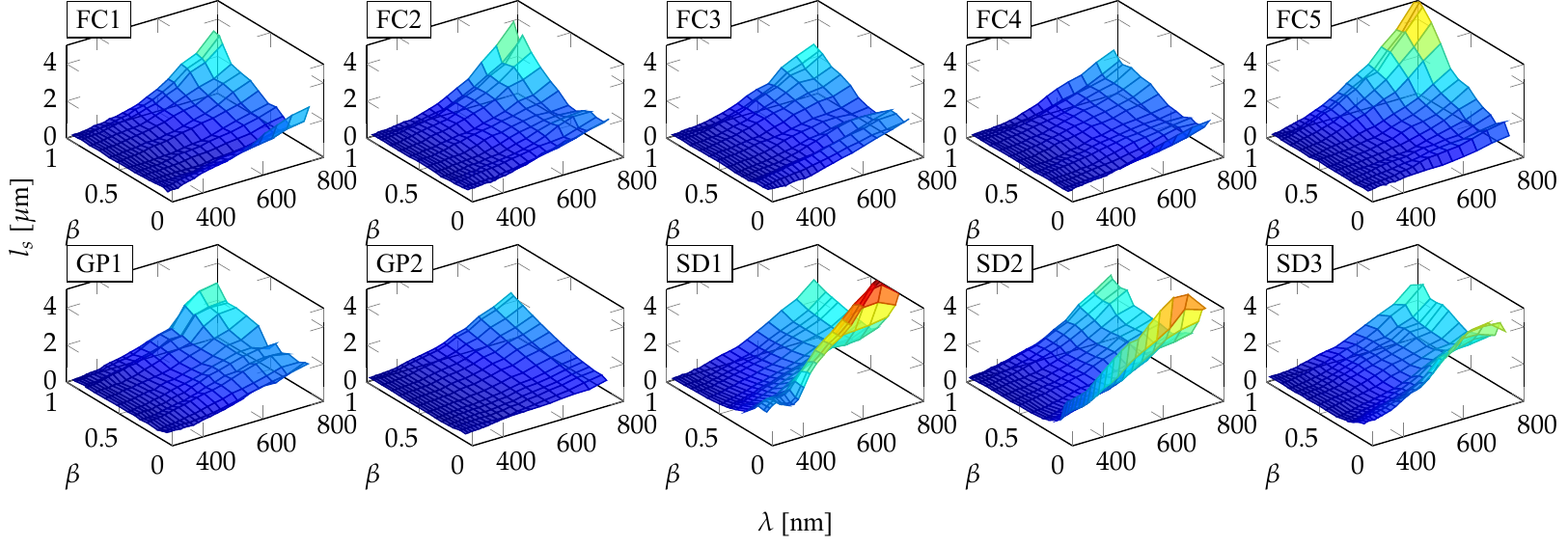}
  \caption{The scattering mean free path, $l_s$, as function of
    wavelength $\lambda$, and anisotropy $\beta$. The plots 
    demonstrate how, especially on large wavelengths,
    the scattering can be increased with anisotropy
    optimisation \label{fig:ls}}
\end{figure}

By replacing the scale parameters and diffusion coefficients in
equations \eqref{eq:GP},\eqref{eq:ch}, and \eqref{eq:fch} with ones
that are independent along different axes we can extending our
simulations to model also anisotropic structures \cite{Essery1991} and
using the Minkowski tensors, we can both induce and quantify
structural anisotropy, and therefore systematically map its effect on
the photonic response. Thus, for all the model types, we created a series of
structures with anisotropy ranging between $\beta \in [0, 1]$, while
keeping the other parameters, $V_0=0.3$ and $l_c=\SI{300}{nm}$
fixed. The values of $\beta$ were roughly evenly spaced since a
structure with an arbitrary anisotropy value has to be iteratively
searched.

The result are shown Figure \ref{fig:aniso}. In all cases the
reflectance shows improvement with increasing anisotropy usually
peaking to $R\approx 0.6$ between $\beta \in [0.4,0.6]$. It is
interesting to note that, thanks to release of the X-ray tomography
data sets of the beetle scales to public domain by Burg and coworkers
\cite{Burg2020}, we were also able to precisely quantify that
that disordered structures inside both \textit{Lepioda stigma} and
\textit{Cyphochilus sp.}~scales are also anisotropic, $\beta=0.7$ and
$\beta=0.6$ respectively.  In the latter case there has been some
speculation whether the structure is isotropic or not, but the
quantitative analysis confirms that the original anisotropy claim
\cite{Burresi2014,Cortese2015} is valid. What is most surprising, for the
simulated structures, is that the system that shows worst reflectivity
in the isotropic case, FC5, reaches the highest reflectivity
$R\approx0.63$ at $\beta \approx0.15$ among all the models, suggesting
that even relatively poor reflectance can be optimised with
anisotropy.

One should note that all the structures, at high anisotropy,
start to resemble 1D multilayer-like systems, and
therefore we would expect them also show high reflectivity 
due to consistency with literature reports
\cite{Bertolotti2005}. In the supporting information
we demonstrate that correlation lengths are merely shifted
and by reoptimisation the higher reflectance levels can be
recovered (cf. Figure S2).

Moreover, by comparing mean free paths as function of
anisotropy, as shown in Figures \ref{fig:ls} (cf. ESI for 
transport mean free path $l_t$) we observed the 
spectrally averaged mean 
free paths are rather similar between the different structures,
and differences are mostly related to spectral
variances. Furthermore these variances are decreased with increasing 
anisotropy with bottle neck like optima between $\beta \sim 0.2$--$0.7$
suggesting anisotropy plays an important role in whiteness optimisation.

\subsection*{Feature analysis}
So far we have mostly related topologically invariant features such
filling fraction, correlation length, and anisotropy to reflectance
properties. Ultimately one aims to arrive at an analytical model for
predicting photonic properties of an arbitrary disorder structure via
examining above mentioned structural properties. Deriving
such model is very difficult task, and instead a black box
approach is needed. Machine Learning (ML) methods have become very 
popular in such problems. In particular deep learning based 
(and similar) ML methods are becoming popular for disorder structures
investigations in photonics and material science\cite{Li2018,Ma2021}, 
and allow the use of raw 3D structures in the learning process. While 
those approaches are in principle very powerful in predicting the 
output (photonic) from the inputs (3D structure), given large enough
training dataset, the challenge of interpreting the structure-property
relationship between the two remains. Given that we have the various
structural descriptors, such as Minkowski functions and anisotropy
values, using them instead of the raw 3D structures for ML regression
analysis, we can quantitatively estimate the importance of each
features, and since each of those features physical interpretation 
also the inference will be more interpretable.

Thus we simply collect all the relevant structural features (a.k.a
labels), for each simulated structure and reflectance spectra 
$R(\lambda)$, and use regression analysis with the following
mapping

\begin{align}
  [l_c,f_{\min},V_0,V_1,V_2,V_3,\beta^{2,0}_{0},\beta^{0,2}_{1},\beta^{2,0}_{1},
  \beta^{2,0}_{2},\beta^{2,0}_{2},\beta^{2,0}_{3},b_0,b_1,b_2] \to R_{\text{avg}}(\lambda) \label{eq:ml}
\end{align}
where $f_{\min}$ is the correlation strength (cf. ESI),
$\beta^{r,s}_\nu$ are the anisotropy factors \cite{Schroeder-Turk2013},
and the Betti numbers $b_0$, $b_1$, $b_2$ measure the number of
particles, loops, and cavities respectively \cite{Pranav2019}.

\begin{figure}[h!]
  \centering
  \includegraphics[width=0.99\textwidth]{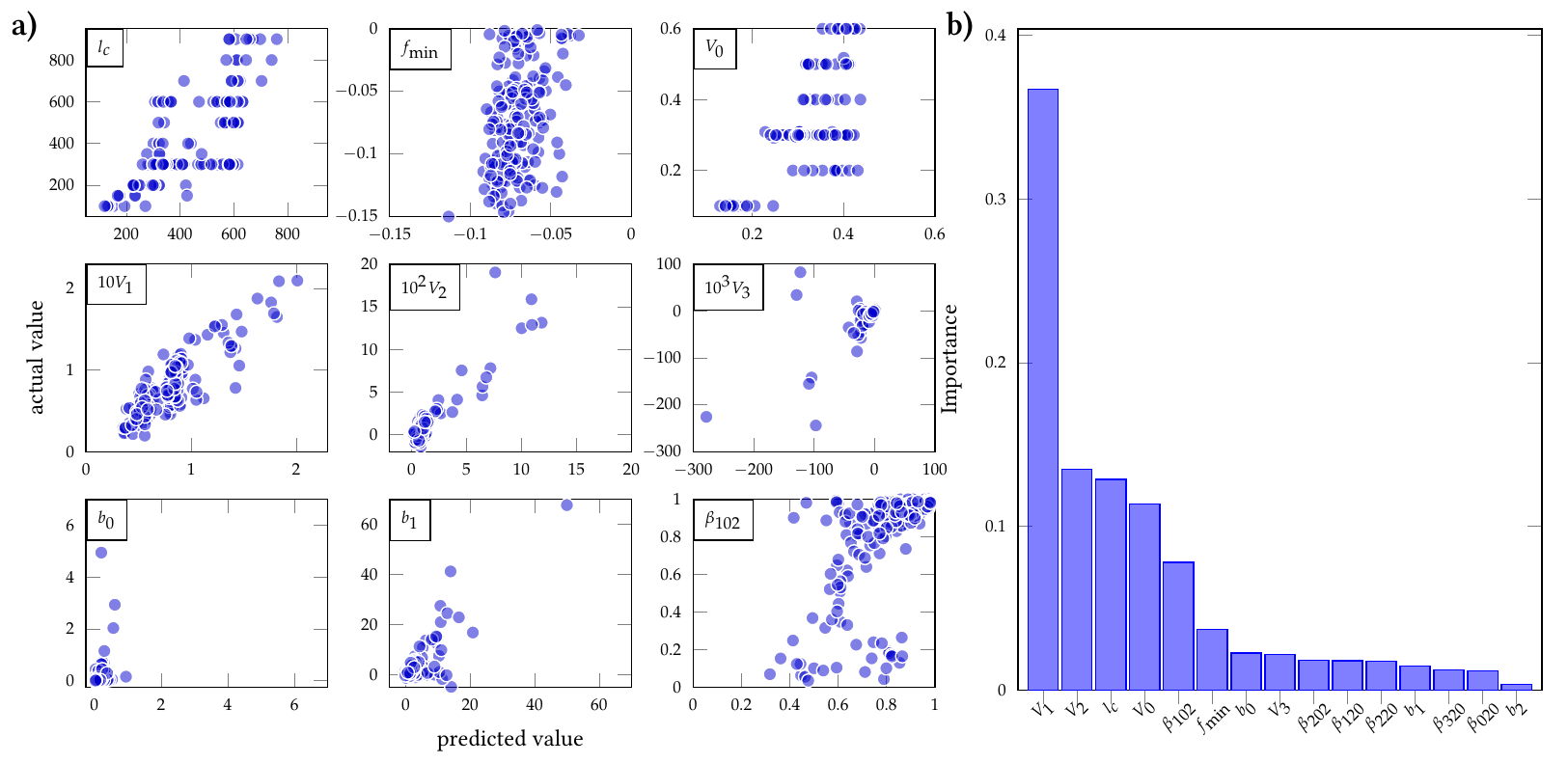}
  \caption{Feature importance analysis using Random Forest
    regression. \textbf{a)} Plots of predicted and actual values of
    the various features (Only the 9 most relevant are shown for space
    considerations). A good correlation between predicted and actual
    value (diagonal distribution of points) is interpreted 
    to signify high relevance to reflectance. \textbf{b)} 
    Quantitative analysis of the feature importance. Value of 1 
    signify high and 0 low importance.  \label{fig:feature}}
\end{figure}

A practical way to assess the importance of each feature is to divide the
data to training and test sets, and compare how accurately a ML model,
trained with former dataset, can predict a particular feature from the
latter. Features that are easy to predict correctly can be interpreted
to have a higher significance for the output, ($R(\lambda)$), and thus
imply a stronger structure-property relationship.

For instance using the popular Random Forest (RF) ML regression model 
for equation \eqref{eq:ml}, we observed in Figure
\ref{fig:feature}a), that most easily predictable features, from
reflectance spectra, are
the surface area $V_1$, integrated mean curvature $V_2$, and
correlation length $l_c$. Furthermore, using quantitative importance 
analysis, we can give percentual estimates of importance of each 15
features to reflectivity, with the five most important being $V_1$
(\SI{37}{\percent}), $V_2$ (\SI{13}{\percent}), $l_c$
(\SI{13}{\percent}), $V_0$ (\SI{11}{\percent}), $\beta$
(\SI{8}{\percent}). These values should however be taken as tentative,
since the importance analysis is rather sensitive parameter
limits. E.g. structures with very low filling fractions, 
$V_0 < \SI{10}{\percent}$,  will have very low reflectance, and 
including them feature analysis will result in higher
weighting of $V_0$ in importance, and the same is true for
$l_c < 100$nm. Therefore we performed the feature analysis 
above these limits, where we still have reasonably amount of 
reflectance. The suggestion that the surface area $V_1$ and 
the mean curvature $V_2$ are the most important features is a consequence
of the fact that the colloidal systems (FC1 \& SD2) seem to 
be most reflective. This indicated that disordered assembly of 
particles would give the highest reflectance. To 
test this hypothesis we performed
series of molecular dynamics particle simulations with the same
correlation lengths $l_c=$\SIrange{100}{900}{nm}. To 
also check if the reflectance is dependent on the
particle shape, we carried out the simulations with both spheres and
tetrahedrons, as the latter has a higher integrated mean curvature,
due to sharp edges, and increased surface area compared to the former.

The results (cf. ESI, Figure S1) indeed show that the spheres 
outperform the other systems in reflectivity, whereas tetrahedrons 
show only moderate reflectance levels, suggesting that the minimal 
surface area with high mean curvature, under the constrain of 
filling fraction and correlation length, is 
a relevant factor for reflectance. Finally, whether spheres 
are, in general, the optimal shape for particle system in respect 
to reflectance is interesting but out of the scope of this work.

\section*{Conclusion}
Using different stochastic models we synthesised, \textit{in silico},
a diverse set of disordered structures varying in shape,
connectiveness, curvature, surface properties, volume, percolation,
number of particles and anisotropy. We then measured their 
reflectance properties, quantified their structural differences, 
and analysed the structure-property relationships. Our extensively 
research revealed that all the investigated disordered systems
exhibit similar unimodal
behaviour in respect to filling fraction and correlations lengths.
By utilising the Minkowski measures we were able for the first time
quantitatively relate structural features, including anisotropy, to
optical properties suggesting that with proper tuning of the second
order statistical features ($l_c$ and $V_0$), and anisotropy brilliant
whiteness can be achieved in any system. While colloidal 
systems have the highest reflectance due  their 
to small surface area $V_1$ and high mean curvature $V_2$, 
due to the lack physical support at optimal filling fraction
$V_0=$\SIrange{30}{40}{\percent}, they are not a feasible solution
in the context of material fabrication. Therefore whiteness
must be realised with continuous structures. However
in this case, we observed that choice of particular morphology becomes
irrelevant as we have demonstrated. From an industrial and design point
of view this is very fortunate, as one is not limited to trying to realise a
particular morphology for optimal whiteness, but instead the efforts
can be focused on optimising the before mentioned invariant parameters. 

The results also suggest that one has to be careful when inferring the
about underlying formation mechanism of biological disordered systems
using optical response and $S_2(r)$ alone, due to non-uniqueness of these
characteristics. Conversely this would also explain why there are many
different disorder system in nature that exhibit whiteness
\cite{Wilts2017,Xie2019,Yu2019}, as the results suggest that any disordered
system generated by a tuneable mechanism can be optimised for
brilliant whiteness.

In conclusion our work suggests that there is no unique route to
brilliant whiteness, and instead it can be realised from many
different starting positions.

\section*{Methods}

All the \textit{in silico} syntheses were carried out in a
$ N \times N \times N$ cubic grid with periodic boundary conditions,
and set to correspond to a $L^3=$\SI{5}{\mu m}$\times$\SI{5}{\mu
  m}$\times$\SI{5}{\mu m} box. A value of $N=100$ was used for all
structures with $l_c \geq \SI{200}{nm}$, and $N=200$ otherwise.

\subsection*{Gaussian processes}
The GP1-2 models where synthesised spectrally using fast Fourier
transform (FFT) techniques
\cite{Robin1993,Ruan1998,Mack2013,Nerini2017}

\begin{align}
  f(\mathbf{x}) = \mathbf{S}*\mathbf{N}= \operatorname{FFT}^{-1}
  \left[
  \operatorname{FFT}(\mathbf{S})\cdotp\operatorname{FFT}(\mathbf{N})
  \right] \label{eq:gp}
\end{align}
where
$\mathbf{S}=[\mathbf{S}_1,\dots,\mathbf{S_N}]$,
$\mathbf{N}=[\mathbf{N}_1,\dots,\mathbf{N}_N]$ are
$(N \times N \times N)$ arrays and

\begin{align*}
  \mathbf{S}_k =
  \begin{bmatrix}
    s_{11k} & s_{12k} & \dots & s_{1Nk} \\
    s_{21k} & s_{22k} & \dots & s_{2Nk}\\
    \vdots & \vdots & \ddots & \vdots \\
    s_{N1k} & s_{N2k} s & \dots & s_{NNk} 
  \end{bmatrix},\ \mathbf{N}_k =  \begin{bmatrix}
    n_{11k} & n_{12k} & \dots & n_{1Nk} \\
    n_{21k} & n_{22k} & \dots & n_{2Nk}\\
    \vdots & \vdots & \ddots & \vdots \\
    n_{N1k} & n_{N2k} & \dots & n_{NNk} 
  \end{bmatrix}, 
\end{align*}
and $s_{ijk} = S_2(r_{ijk})$,
$r^2 = (\tfrac{N}{2} - i)^2 + (\tfrac{N}{2} - j)^2 + (\tfrac{N}{2} - k)^2$,
$n_{ijk} \sim \mathcal{N}(0,1)$

\subsection*{Phase-field simulations}
The Cahn-Hilliard model used for generating the spinodal decomposition 
structures SD1-3 is given by

\begin{align}
  \frac{\partial f}{\partial t} &= \nabla^2 
  M\left[\frac{\partial W(f)}{ \partial f} - \tfrac{1}{2} \epsilon (\nabla f)^2 \right] \label{eq:ch} \\
    W(f) &= A f^2(1-f)^2 
\end{align}
where $M$ and $\epsilon$ are the mobility and the diffusion constants,
and $W(f)$ is the mixing energy between the two equilibrium phases, and
was numerically solved using semi-explicit spectral 
method\cite{Chen1998,Biner2017}

\begin{align}
  \mathbf{F}^{[n+1]} = \frac{\mathbf{F}^{[n]}
  - \Delta t k^2 M W'(\mathbf{F}^{[n]})  
  }{1 + \Delta t k^4 M \epsilon}
\end{align}
where $\mathbf{F}^{[i]} = \operatorname{FFT}(\mathbf{f}^{[i]}), $with time step
$ \Delta t = 1 \times 10^{-3}$, $A = 1$, $M = 1$, and the coefficient
$\epsilon$ was chosen from [0.04,1.00] to reach the desired length
scale.

The FC1-5 models were created with Functionalised Cahn-Hilliard model \cite{Gavish2012,Jones2012}

\begin{align}
  \frac{\partial f}{\partial t} &= \Delta 
  \left[  E_b
     \left(
       \epsilon^2 \Delta - W''_b(f)
     \right)
     \left(
       \epsilon \Delta f - W'_b(f)
     \right) + \eta_h \epsilon^2 \Delta f - \eta_m W'_s(f)
  \right] \label{eq:fch} \\
 W_s(f) &= W_b(f) = \frac{1}{2}   \left(f + 1  \right)^2 \left( \tfrac{1}{2}
  (f - 1)^2 + \frac{\tau}{3}(f -2)  \right).
\end{align}
using the CUDA code of Jones \cite{Jones2013} with $\eta_h=5$ and
$\eta_m=[-7.25,-0.5,3,6,10]$ for FC1-5 respectively.

\subsection*{Synthesis of anisotropic structures}
Anisotropic structures where synthesis using a the method of Essery
\cite{Essery1991} where the mobility coefficient $\epsilon$ in equations
\eqref{eq:ch}, and \eqref{eq:fch} was replaced by tensor

\begin{align}
  \epsilon =
  \begin{bmatrix}
    \epsilon_x & 0 & 0 \\
    0 & \epsilon_y & 0  \\
    0 & 0 & \epsilon_z
  \end{bmatrix}
\end{align}
and a similar method for the GP models by anisotropic scaling
$r^2 = \epsilon^2_x(\tfrac{N}{2} - i)^2 + \epsilon^2_y(\tfrac{N}{2} -
j)^2 + \epsilon^2_z(\tfrac{N}{2} - k)^2$ in equation \eqref{eq:gp}.

\subsection*{Conversion to binary fields}
The stochastic fields $f(\mathbf{x}):\mathbb{R}^3 \to \mathbb{R}$
synthesised with the different methods were converted to the two phase
disordered structures $I(\mathbf{x}):\mathbb{R}^3 \to \{0,1\}$ using a
simple thresholding scheme

\begin{align*}
  I(\mathbf{x})=
  \begin{cases}
    0, & \text{if } f(\mathbf{x}) < p_0 \\
    1, & \text{else }  
  \end{cases}
\end{align*}
where $I(\mathbf{x})$ is an indicator function , and $0$ and $1$
represent the empty and the solid phase, respectively, and $p_0$ is the
threshold value. Thus the final filling fraction
$V_0=\langle I(\mathbf{x}) \rangle$ of structures is determined by the
choice of $p_0$. While such systems might be physically difficult to
realise, we are primarily interested in understanding the optical
properties of disorder systems, and question about chemical synthesis
of potential structures are beyond the scope of this paper.

For the FDTD simulation, the structures $f(\mathbf{x})$ were  imported to
USCF chimera \cite{Pettersen2004} and converted to STL stereolitography 
files using a similar level set scheme.

\subsection*{Particle simulations}
The additional particle simulations (see ESI) were conducted with
HOOMD-blue\cite{Anderson2016,Anderson2020}package using molecular
dynamics simulation with Langevin integrator and Weeks-Chandler-Andersen
potential for spheres, and hard particle Monte Carlo simulation for
the tetrahedrons.

\subsection*{Calculation of Minkowski measures}
The Minkowski functionals of $I(\mathbf{x})$

\begin{align}
  \begin{array}{ll}
  V_0(I) = \int_I dV, &  V_1 (I) =  \frac{1}{3}\int_{\partial I} ds   \\
  V_2(I) = \frac{1}{6}\int_{\partial I} \frac{1}{2} 
           \left(
           \frac{1}{r_1(s)} + \frac{1}{r_2(s)}
           \right) ds, &
  V_3(I) = \frac{1}{3} \int_{\partial I} \frac{1}{r_1(s)r_2(s)} ds
  \end{array}
\end{align}
where $r_1(s)$ and $r_2(s)$ are the maximum and minimum curvature
radii \cite{Arns2010}, the Minkowski tensors $W^{r,s}_{\nu}$ and
anisotropy factors

\begin{align}
  \beta^{r,s}_{\nu}:=\frac{|\mu_{\min{}}|}{|\mu_{\max{}}|} \in [0,1],
\end{align}
where $\mu_{\min}$ and $\mu_{\max}$ are the minimal maximal
eigenvalue of $W^{r,s}_{\nu}$, were calculated using the Karambola 
software package \cite{Schroeder-Turk2011,Schroeder-Turk2013} 
from the STL converted files. Anisotropy values for beetle scales were 
calculated from the files \texttt{CY\_cube.npy} and
\texttt{LS\_cube.npy} of Burg and coworkers' dataset\cite{Burg2020}.

\subsection*{Calculation of two-point correlation functions}
To determine the characteristic length scale of the simulated
structures $I(\mathbf{x})$, we used radially averaged 2-point
correlation function calculated using FFT method\cite{Jiao2007}

\begin{align}
  S_2(r) =  \frac{\sum_{l,m,n \in \Omega} \operatorname{FFT}^{-1}
  \left(
  |\operatorname{FFT}(I(\mathbf{x}))|^{2}
  \right)}{\omega}
\end{align}
where
$\Omega = \left\{(l,m,\ n) \ | \ l^2 + m^2 + n^2 = r^2,\ r \leq N/2
\right\} $ and $\omega$ is the number of elements in $\Omega$. The
correlation length, $l_c$, was defined to be the distance where
2-point correlation function has its first minima
$l_c := {\mathrm{argmin}}_r\ S_2(r)$. For cases where $S_2(r)$ was
monotonically decreasing we used definition of $l_c$ as the minimum
distance $r_n$ where relative change,
$\frac{S_2(r_n) - S_2(r_{n+1})}{S_2(0)}$ , was less than
$\SI{0.1}{\percent}$. Correlation strength was defined as the relative
depth of the minima of $S_2(r)$

\begin{align}
  f_{\min} = \min{
  \left(
  \frac{S_2(r) - S^2_2(0)}{S_2(0) - S_2^2(0)}
  \right)}
\end{align}

\subsection*{Finite-difference time domain (FDTD) calculations}
The FDTD simulations were carried out using Lumerical 2020a-r5 (Ansys
Canada Ltd), with periodic boundary conditions and perfect matching
layer boundaries in x, y-directions and broad band source
$\lambda \in [300,800]$nm in p-polarisation (TM-mode) coming from
vertical direction. The refractive index for the material 
was set to $n=1.50$ in all cases. The numerical stability and 
convergence was ensured with the adequate boundary condition and the
simulations were carried out until all incoming light had either 
reflected or transmitted. For the mean free path calculations, 
the ballistic transmission was recorded using additional TM and 
TE monitors.

\subsection*{Regression analysis}
The feature importance calculations were carried out using
\textit{RandomForestRegressor} of Scikit-learn\cite{Pedregosa2011} 
Python library with 200 trees for features collected from 400 
simulated  structures.

\subsection*{High performance computations}
The FDTD simulations in this work were performed using resources
provided by the Cambridge Service for Data Driven Discovery (CSD3)
operated by the University of Cambridge Research Computing Service
(\url{www.csd3.cam.ac.uk}), provided by Dell EMC and Intel using
Tier-2 funding from the Engineering and Physical Sciences Research
Council (capital grant EP/P020259/1), and DiRAC funding from the
Science and Technology Facilities Council (\url{www.dirac.ac.uk}).\\

\noindent The in silico synthesis of the FC1-5 structures were 
performed using computer resources provided by the 
Aalto University School of Science
''Science-IT'' project (\url{https://scicomp.aalto.fi/}).

\section*{Code and data availability}
Matlab code for generating the GP1-2 and SD1-3 structures 
and $S_2(r)$ calculations, and additional data related to 
this publication is available at the University of Cambridge 
data repository 
(\url{https://doi.org/10.17863/CAM.71288}).

\section*{Acknowledgements}
The authors thank Prof. R\'{e}mi Carminati 
for his helpful suggestions on the manuscript.

J.S.H. is grateful for financial support from the Emil
Aaltonen Foundation. L.S. acknowledges the 
support of the Isaac Newton Trust and the Swiss 
National Science Foundation under project 40B1-0\_198708.  
This work is part of a project that has received
funding from the European Union's Horizon 2020 research and innovation
programme under the Marie Skłodowska-Curie grant agreement No.~893136
and the ERC SeSaME ERC-2014-STG H2020 639088.

\section*{Author contributions statement}
J.S.H did the in silico synthesis and data analysis, G. J and J.S.H
carried out the FDTD simulations, J.S.H, G.J, L.S, and S.V designed
the experiments, commented on results and wrote the manuscript.

\section*{Competing interests}
The authors declare no competing interests.

\section*{Additional information}
Correspondence should be addressed to J.S.H or S.V.

\bibliography{main_ax}

\end{document}


\maketitle

\section{Particle Simulations}
\begin{figure}[h!]
  \centering
  \includegraphics[width=\textwidth]{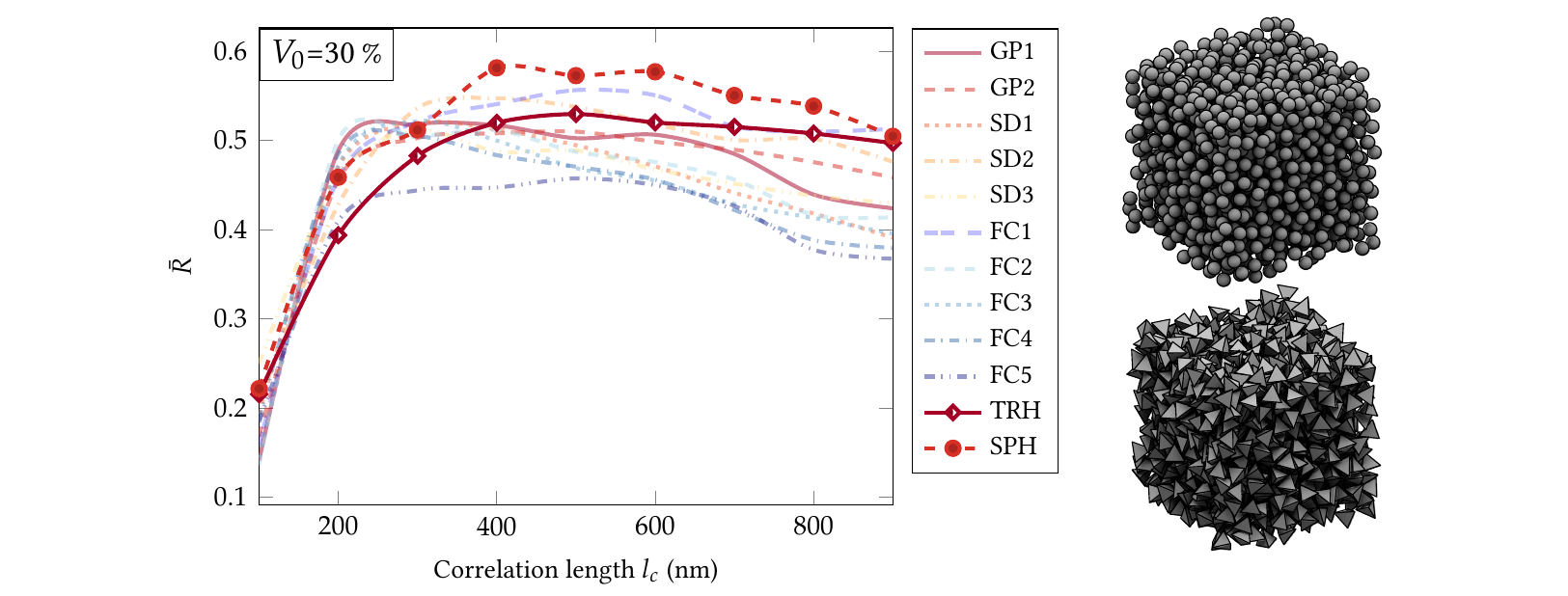}
  \caption{(Left): Comparison of total reflectance of low (spheres -
    SPH) and high (tetrahedrons - THR) mean curvature disordered
    particle systems with the field simulated structures. (Right):
    Examples of disordered spherical and tetrahedron systems.
  \label{fig:MD}}
\end{figure}

To test hypothesis that the average reflectance is dependent on
integrated mean curvature, $V_2$ we carried out molecular dynamics
(MD) simulations to create disordered assembly of spheres and
tetrahedrons with filling fraction of \SI{30}{\percent} and
correlation length ranging from \SIrange{100}{900}{nm},
cf. Fig. \ref{fig:MD}. Spheres, in agreement with other resutls, show
the highest reflectance, where as tetrahedrons only give mediocre
average reflectances. However this in conflict with the suggestion
that the reflectance is only depent on curvature, since tetrahedrons
have in fact a higher $V_1$ due to sharp edges, and also increased
surface area, $V_1$. Thus it seems to be that most apparent
explanation for the high reflectivity of spherical systems are the
high curvature and minimised surface area.

\section{Reflectance Optimization of Anisotropic Systems}
\begin{figure}[h!]
  \centering
    \includegraphics[width=\textwidth]{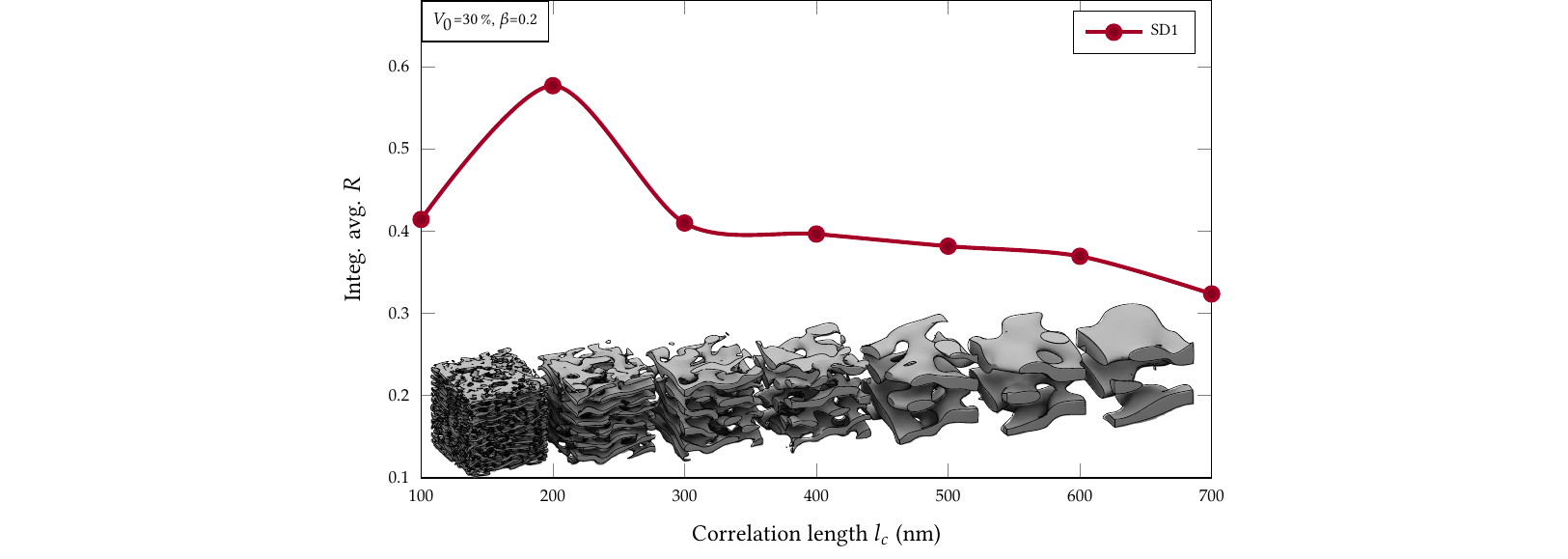}
  \caption{Example of reflectance optimisation of highly anisotropic 
  SD1 by tuning the correlation length.}
  \label{fig:PS1}
\end{figure}

\begin{figure}[h!]
  \centering
  \includegraphics[height=0.85\textheight]{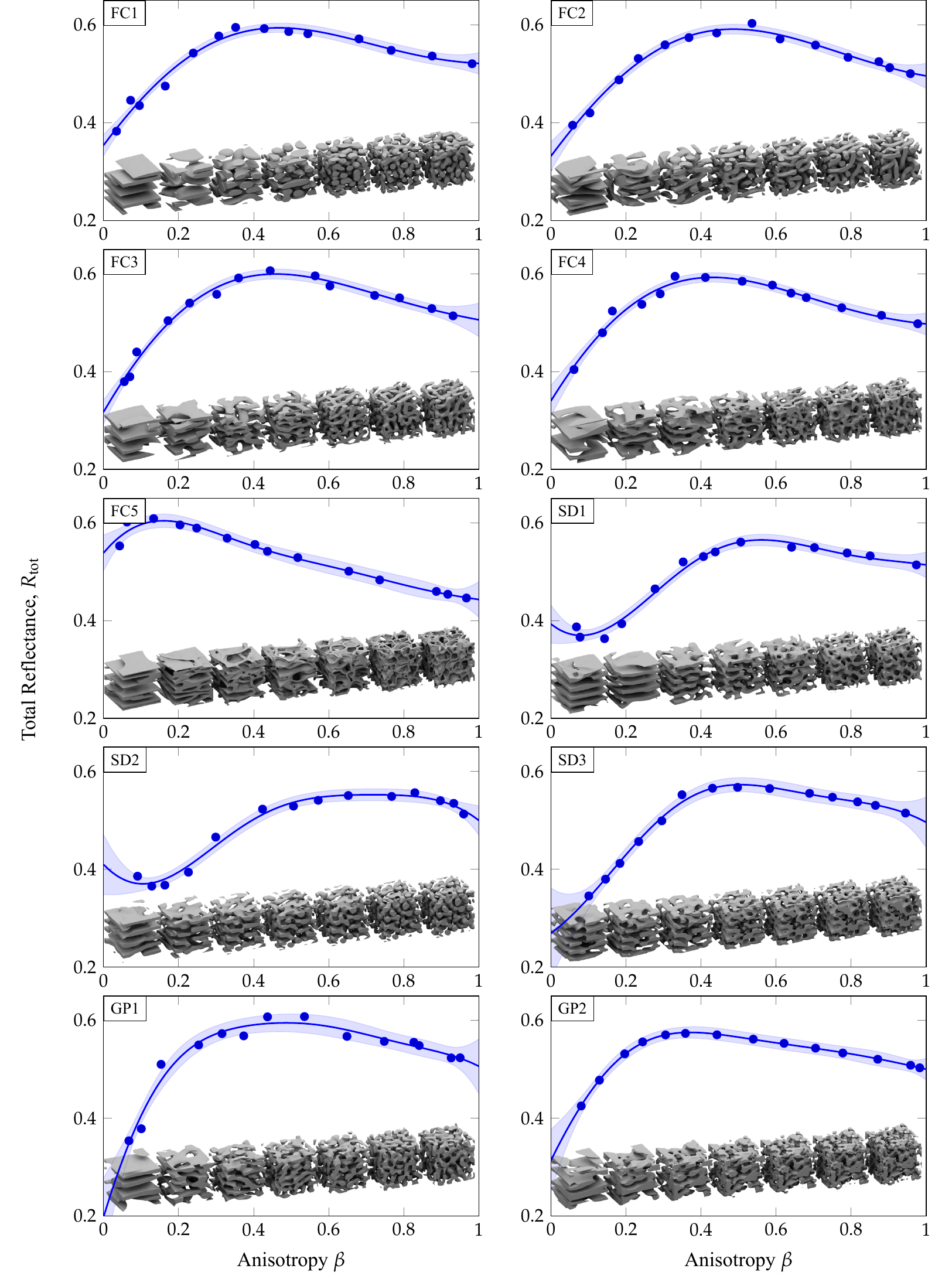}
  \caption{Effect of anisotropy $\beta$ on average
    reflectance. A value of  $\beta=1$ indicates complete
    structural isotropy and $\beta=0$ full structural anisotropy.
    The dotted points represent measured (FDTD simulated) reflectances, 
    and the solid line present Gaussian Process fits and the
    shaded region present estimated confidence intervals.
    \label{fig:aniso}}
\end{figure}

Anisotropy investigations for all the different morphologies is 
shown in figure Figure \ref{fig:aniso}. As noted in the main text 
the reflectance of FC5 system increases as the systems becomes more
anisotropic $\beta \to 0$, while the other systems show decrease 
after initial improvement. For instance the SD1
systems visually start to resemble a perforated multilayer at high
anisotropy values, and one would expect this would lead to higher,
instead of lower, reflectivity. A simple line scan
(cf. Fig. \ref{fig:PS1}) along different correlation length at
$V_0=\SI{30}{\percent}$, $\beta=0.2$ shows that $l_c$ is shifted from
\SI{300}{nm} to \SI{200}{nm}

\newpage
\section{Mean Free Paths}
\begin{figure}[h!]
  \centering
    \includegraphics[width=\textwidth]{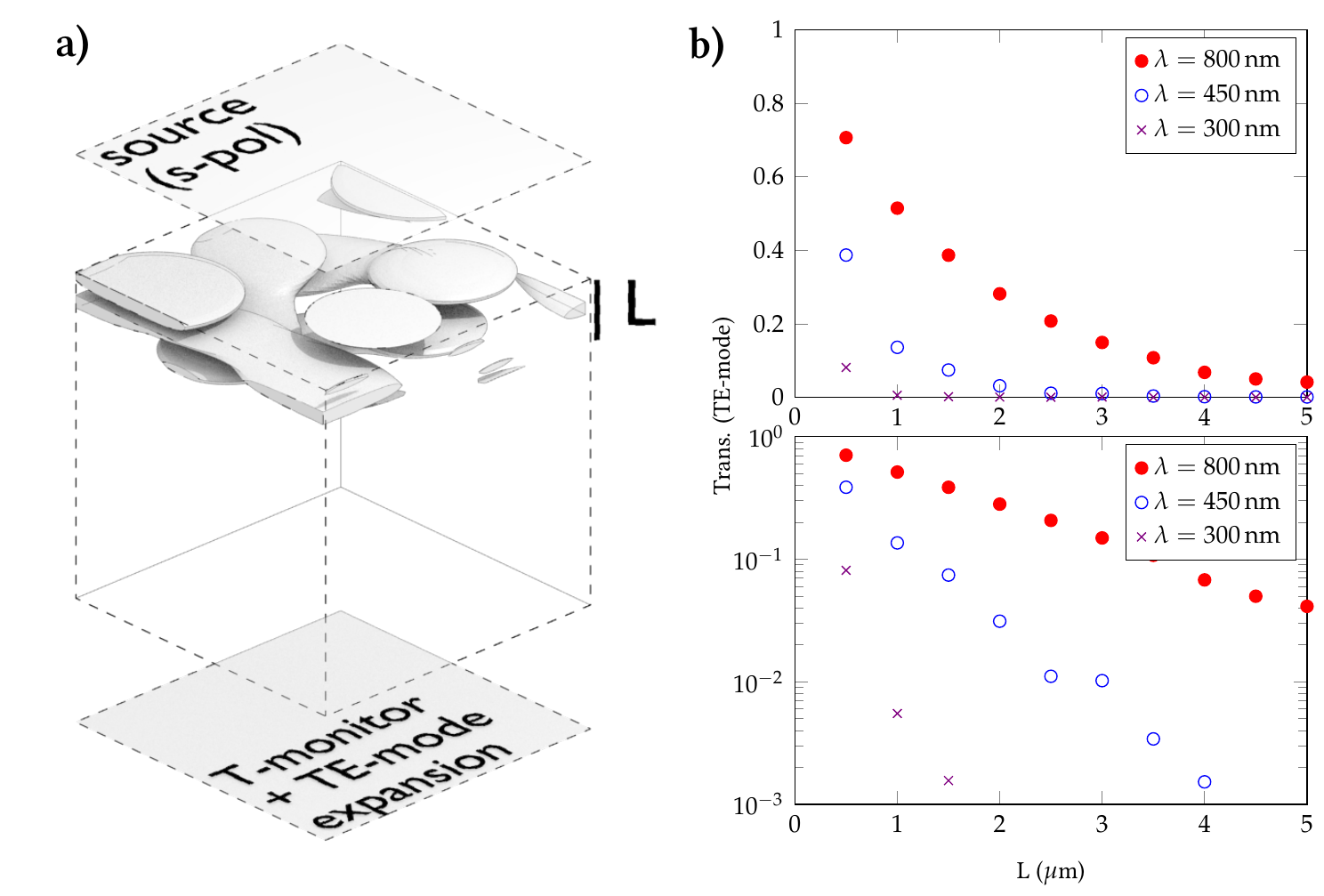}
    \caption{\textbf{a)} Simulation setup for the mean free path
      calculations. The sample thickness was varied by cutting whole
      $\SI{5}{\micro m} \times \SI{5}{\micro m} \times \SI{5}{\micro
        m}$ volume to different thicknesses $L$ and measuring both the
      total and ballistic transmission. \textbf{b)} The results show
      good exponential decay as expected by the Lam-Beer law. }
  \label{fig:fdtd}
\end{figure}
To investigate the topological invariance of the reflectance we varied
the sample thickness $L$ between $\SIrange{0.5}{5}{\micro m}$ in
$\SI{0.5}{\micro m}$ steps to calculate the scattering mean free path
$l_s$ from Lam-Beer law. (top): lin-lin and (bot) log-lin plots.

\begin{align}
  T_{b} = e^{-L/l_s}
\end{align}
and transport mean free path $l_t$ from \cite{Garcia2008}

\begin{align}
  T_{\text{tot.}} &= \frac{2z_0}{L + 2z_0}, \\
  z_0 &= \frac{2}{3} l_t \frac{1 + \mathcal{R}}{1 - \mathcal{R}}
\end{align}
where $T_{b}$ and $T_{\text{tot}}$ are the ballistic and total
transmission respectively, and the coefficient $\mathcal{R}=0.23$ was
calculated using Maxwell-Garnett approximation. The ballistic
transmission was measured using mode expansion monitor in Lumerical.

The mean free paths where calculated using the least squares slope
solution \cite{Bolstad2007} 

\begin{align}
l_{[*]} = \frac{\overline{xy} - \bar{x}\bar{y}}{\overline{x^2} - \bar{x}^2}
\end{align}
where are $x=-\log(T_b)$, $y=-L$ and
$x=2\frac{2}{3}
\frac{1+\mathcal{R}}{1-\mathcal{R}}\frac{1-T_{\text{tot}}}{T_{\text{tot}}}$,
$y=L$ for $l_s$ and $l_t$ respectively.

\begin{figure}[b!]
  \centering
  \includegraphics[width=0.99\textwidth]{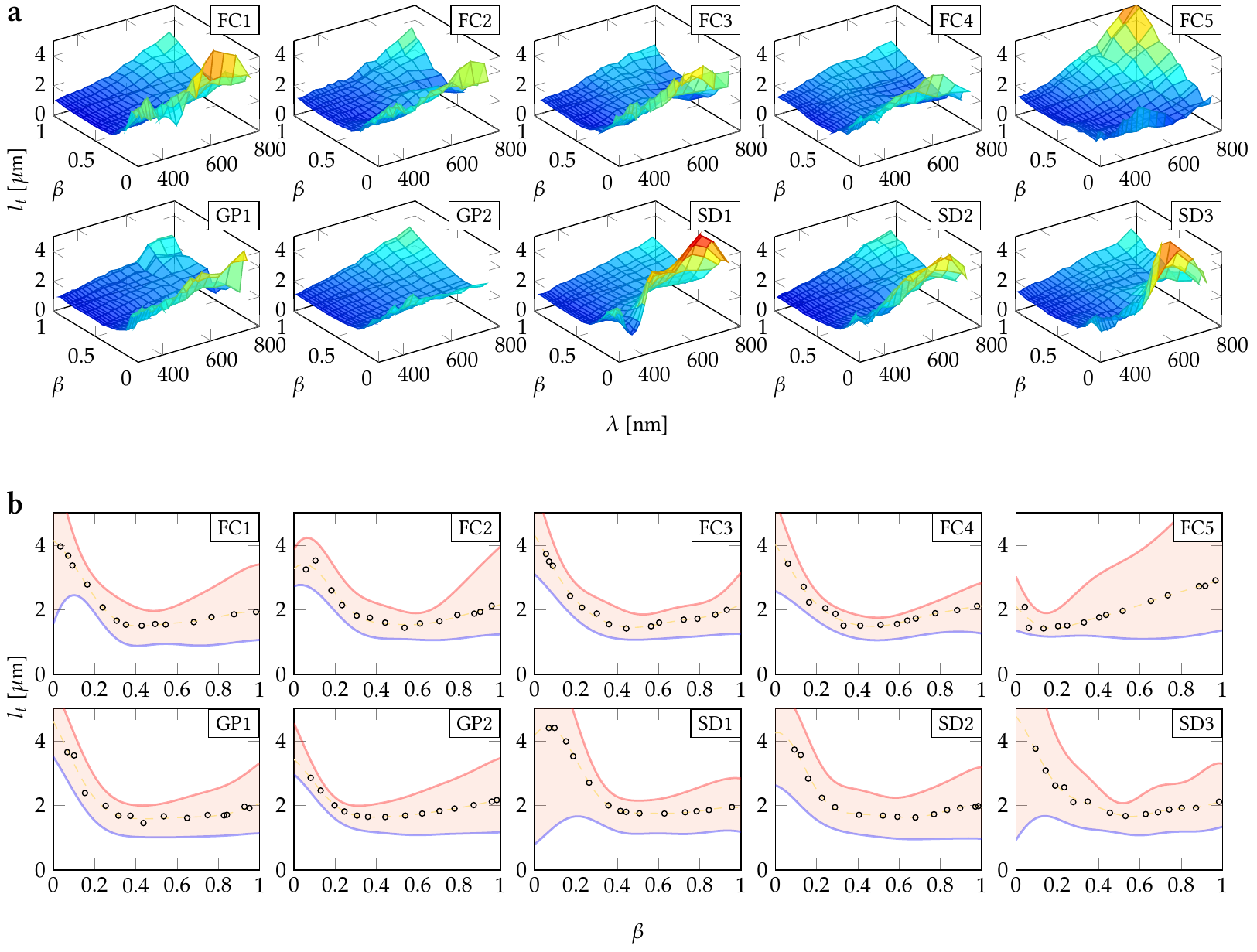}
  \caption{The transport mean free path, $l_t$, as function of
    wavelength $\lambda$, and anisotropy $\beta$. \textbf{a)} 3D plots
    of $l_t$ vs $\beta,\lambda$ and \textbf{b)} marginal plots where
    the shaded area presents variation between spectral extremes (red
    curve for $\lambda=\SI{800}{nm}$, and blue for
    $\lambda=\SI{300}{nm}$) and the markers indicate spectral
    averages. The plots demonstrate how, especially on large
    wavelengths, the scattering can be increased with anisotropy
    optimisation \label{fig:lt}}
\end{figure}

\begin{figure}[b!]
  \centering
  \includegraphics[width=0.99\textwidth]{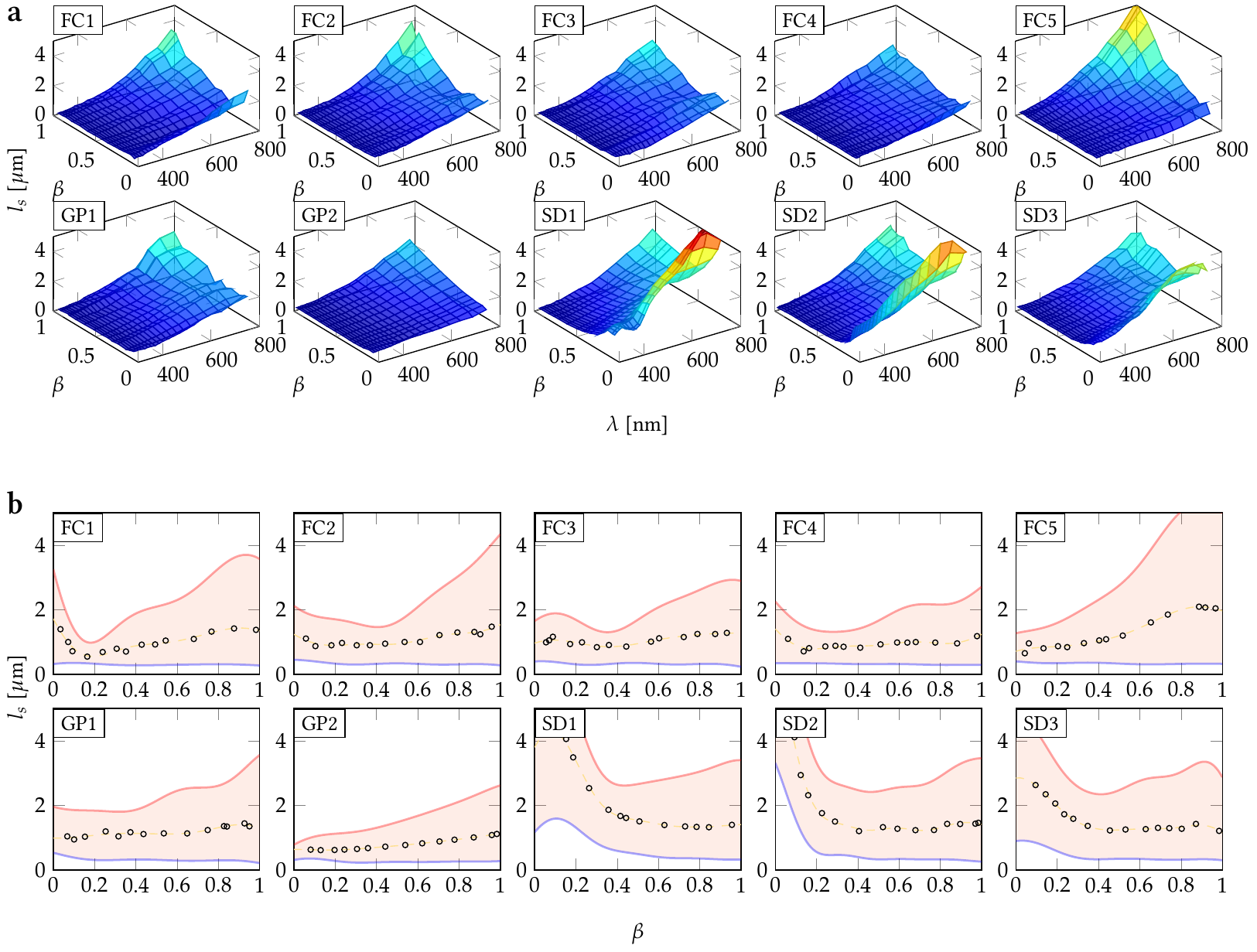}
  \caption{The scattering mean free path, $l_s$, as function of
    wavelength $\lambda$, and anisotropy $\beta$. \textbf{a)} 3D plots
    of $l_s$ vs $\beta,\lambda$ and \textbf{b)} marginal plots where
    the shaded area presents variation between spectral extremes (red
    curve for $\lambda=\SI{800}{nm}$, and blue for
    $\lambda=\SI{300}{nm}$) and the markers indicate spectral
    averages. The plots demonstrate how, especially on large
    wavelengths, the scattering can be increased with anisotropy
    optimisation \label{fig:ls}}
\end{figure}

\clearpage
\newpage
\bibliography{main_ax}